\documentclass{elsart}

\usepackage{natbib}
\usepackage{graphicx}

\usepackage{amssymb}
\usepackage{amsmath}
\usepackage{bm}

\def\url#1{{\ttfamily\def\/{/\discretionary{}{}{}}#1}}
\def\bibcode#1{}

\begin{document}

\begin{frontmatter}
\title{Finding Clusters in SZ Surveys}
\author[address1]{Chris Vale\thanksref{cvemail}},
\author[address1,address2]{Martin White\thanksref{mwemail}}
\address[address1]{Department of Physics, University of California,
Berkeley, CA, 94720}
\address[address2]{Department of Astronomy, University of California,
Berkeley, CA, 94720}
\thanks[cvemail]{E-mail: cvale@astro.berkeley.edu}
\thanks[mwemail]{E-mail: mwhite@astro.berkeley.edu}

\begin{abstract} 
We use simulated maps to investigate the ability of high resolution, low 
noise surveys of the CMB to create catalogues of Clusters of galaxies by 
detecting the characteristic signature imprinted by the Sunyaev Zeldovich 
effect.  We compute the completeness of the catalogues in our simulations 
for several survey strategies, and evaluate the relative merit of some 
Fourier and wavelet based filtering techniques.
\end{abstract}

\end{frontmatter}

\section{Introduction}

Future measurement of the distribution and number density of clusters of 
galaxies will place increasingly important constraints on the nature 
of the universe we live in \cite[e.g. ][]{Bahcall99,RBN02,Voit04}, 
and is a major science goal of upcoming surveys such as 
SZA\footnote{http://astro.uchicago.edu/sza/},
APEX-SZ\footnote{http://bolo.berkeley.edu/apexsz/},
the South Pole Telescope (SPT\footnote{http://astro.uchicago.edu/spt/})
and the Atacama Cosmology Telescope
(ACT\footnote{http://www.hep.upenn.edu/$\sim$angelica/act/act.html})  
These surveys will map the millimeter and 
sub-millimeter sky with unprecedented power and resolution, which will 
enable the construction of a catalogue of clusters detected through the 
thermal Sunyaev Zel'dovich effect (SZE) 
\citep[][for recent reviews see \citealt{Reph95,Birk99,Carl02}]{SZ72,SZ80}
In this paper, we examine different survey strategies 
and signal processing methodologies to enhance this effort.

The imprint of the SZE on the cosmic microwave background (CMB) is 
an integrated effect from the time of last scattering to the present era, 
and as such the SZE signal suffers from projection effects due to other 
objects along the line of sight.  This introduces non-linear complications 
to our signal processing efforts, and makes it impossible to conclusively 
determine the best method by analytic means alone.  We therefore test 
and compare three promising filtering techniques: discrete wavelets, 
continuous wavelets, and Fourier methods.  These are applied to mock SZ maps 
for several different survey strategies, and results for the different 
strategies and filters are computed.

The mock SZ maps are created using an N-body simulation of sufficient volume 
to be a fair sample of the universe.  Due to the current uncertainty in 
both the magnitude of the SZE and of relevant astrophysical foregrounds, 
a detailed modeling of the signal and noise is not currently possible.  
However, some of the complications that will be encountered by actual 
surveys, such as confusion due to projection effects, irregularly shaped 
sources, maps with edges and holes, and spatially varying noise, are 
included in our tests prospective filters.

The outline of the paper is as follows.  We describe our simulations in 
Section \ref{sec:simulations} and our filtering schemes in 
Section \ref{sec:wavelets}.  We then present our results in the context of 
various survey scenarios and signal processing techniques in 
Section \ref{sec:results}, and discuss our conclusions in 
Section \ref{sec:conclusions}.

\section{Simulating the SZE} \label {sec:simulations}

\begin{figure}
\begin{center}
{\includegraphics*[height=6.8cm,width=6.8cm]{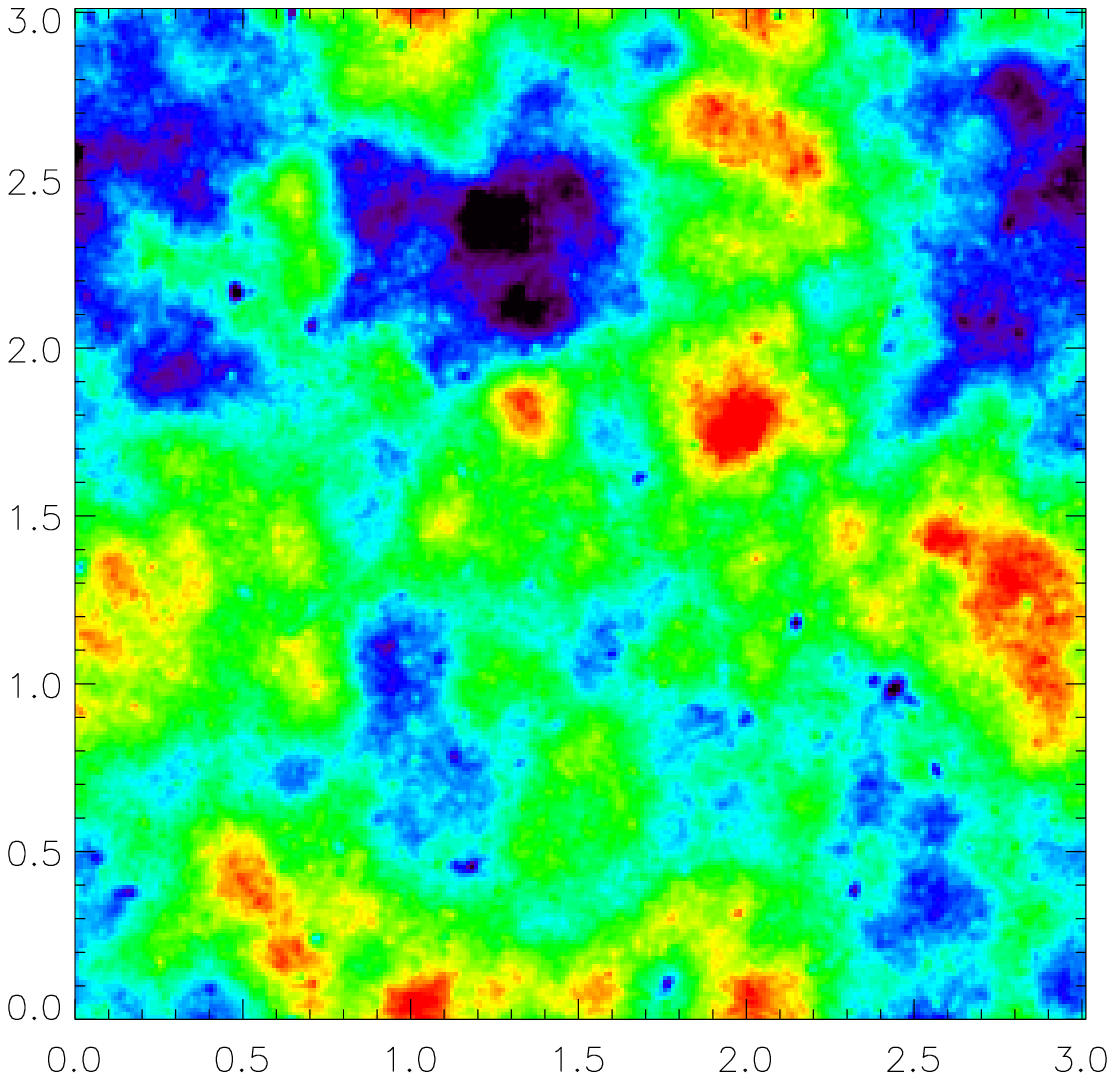}
 \includegraphics*[height=6.8cm,width=6.8cm]{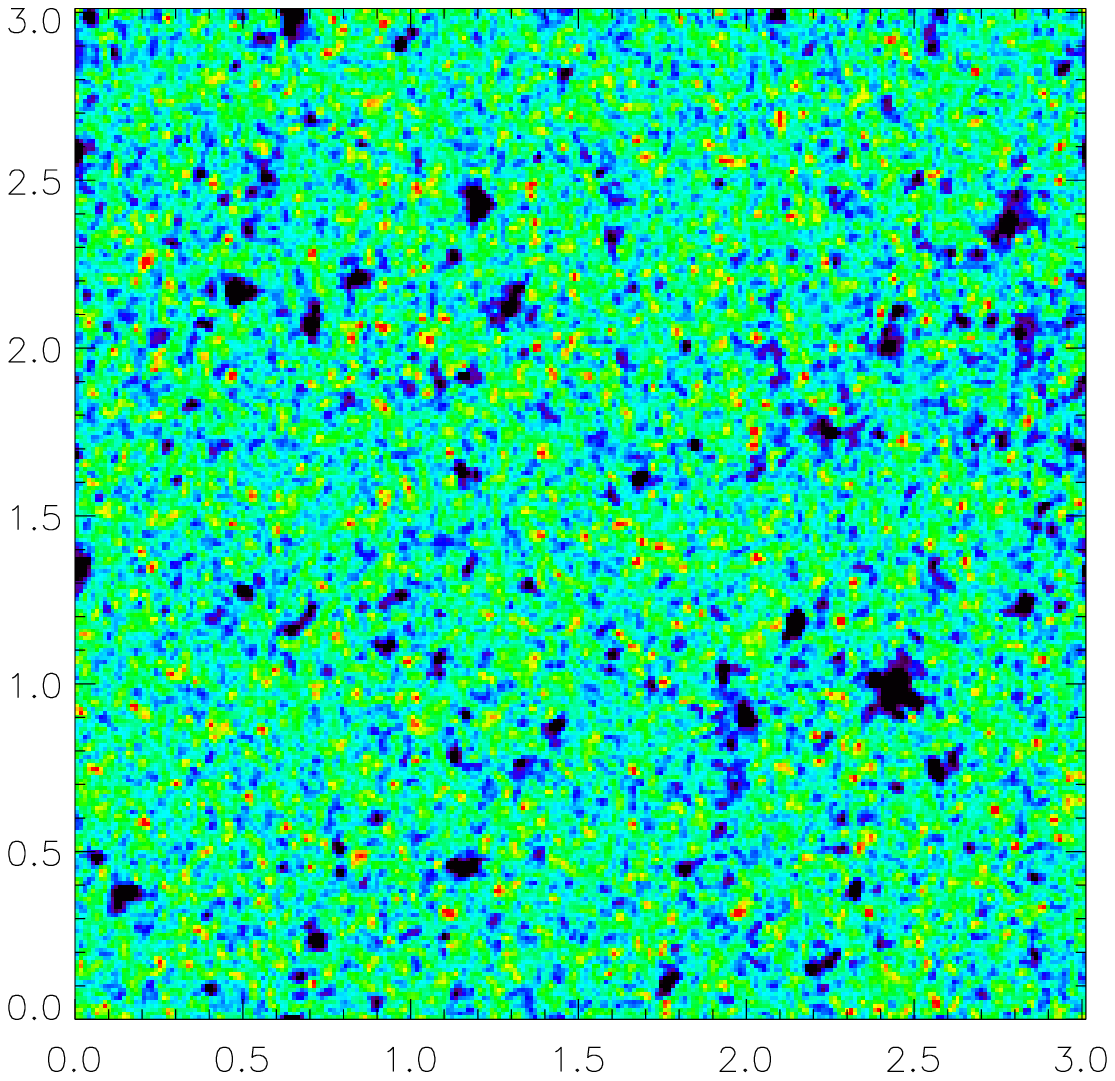}}
{\includegraphics*[height=6.8cm,width=6.8cm]{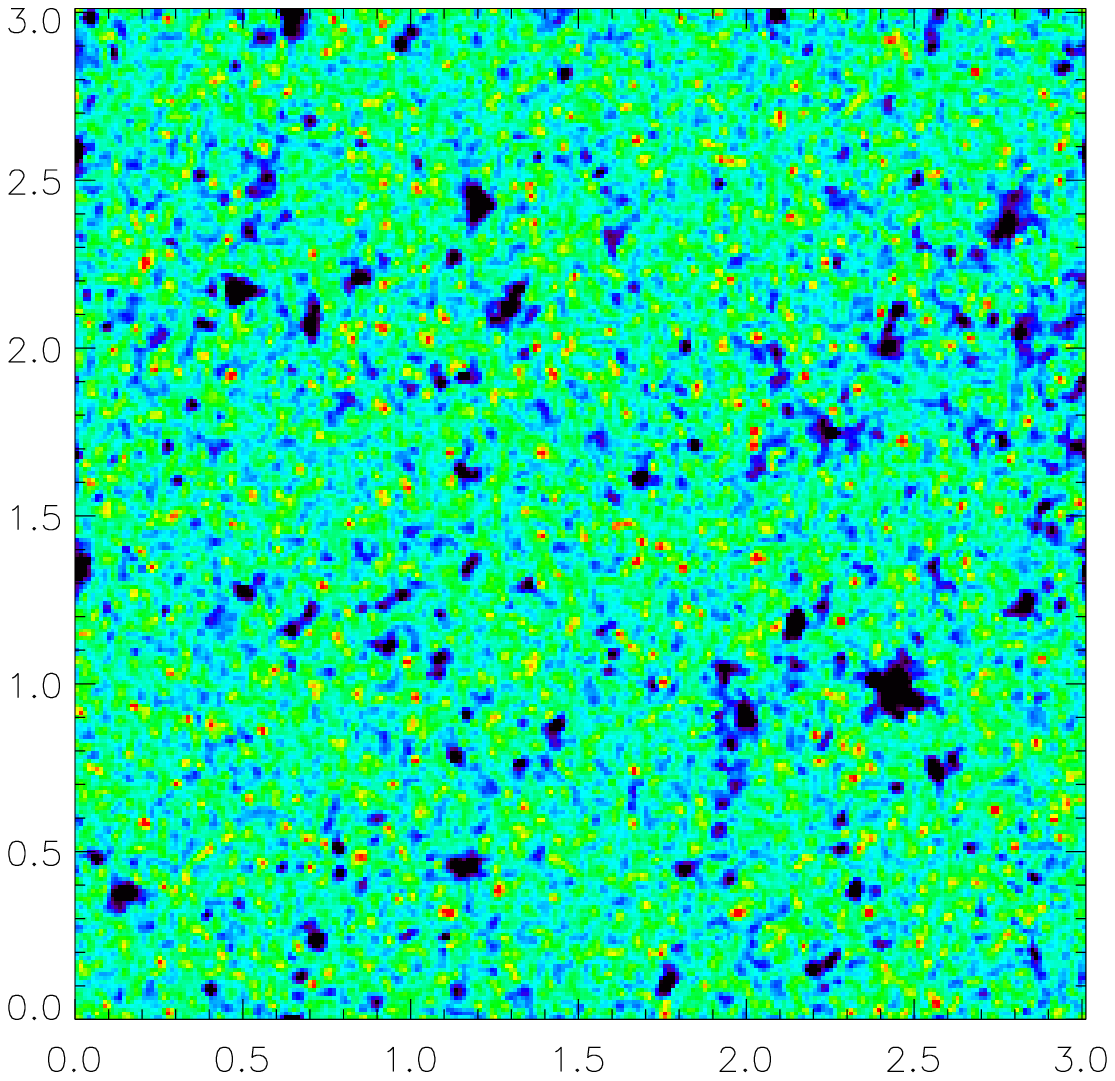}
 \includegraphics*[height=6.8cm,width=6.8cm]{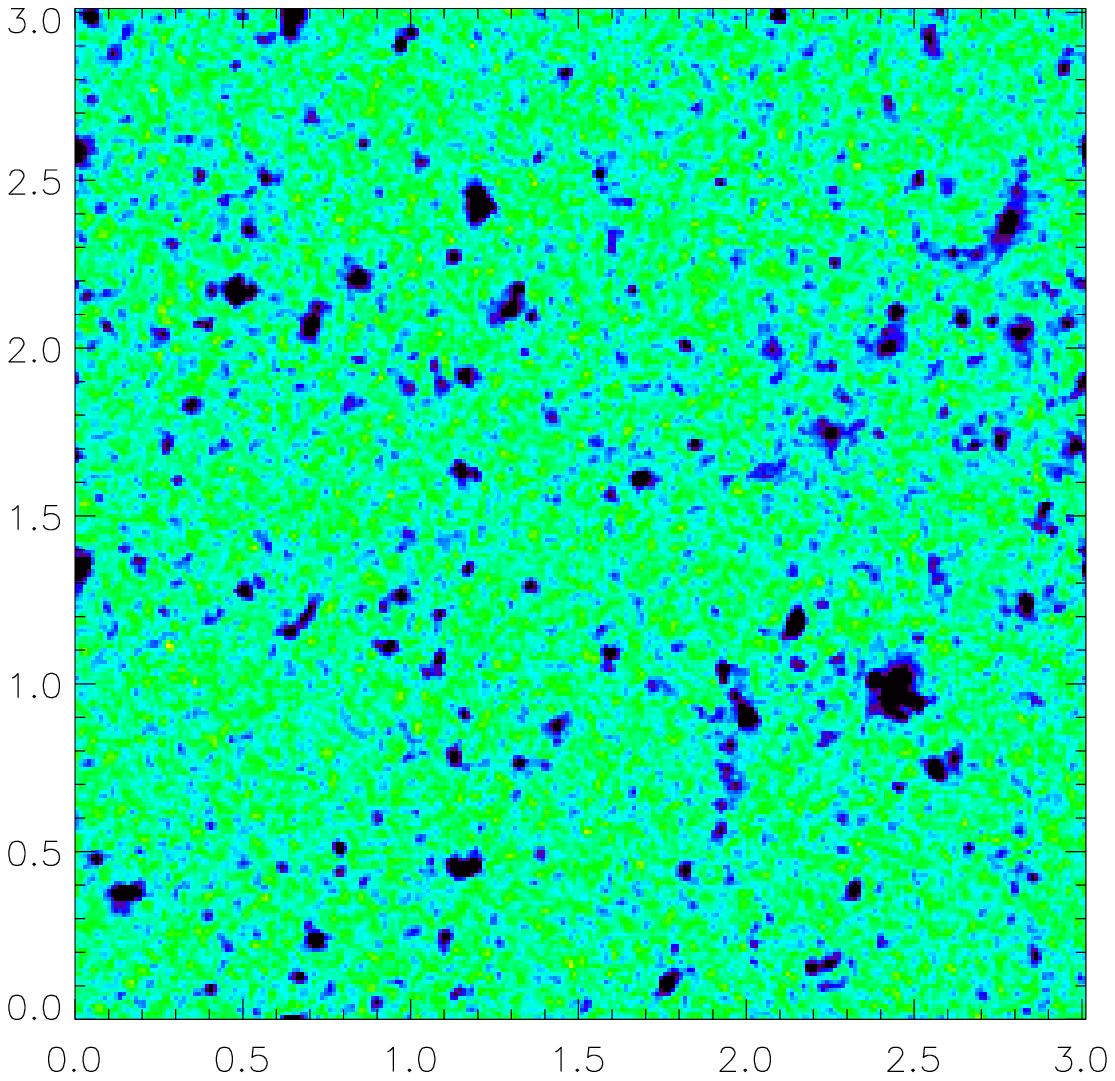}}
\end{center}
\caption{An example of the maps before filtering.  The full map (top left) 
includes the SZE signal (which shows up as cold spots on the map) and all 
sources of ``noise''.  Since the signal is overwhelmed by the primary CMB on 
the angular scale shown here, we display the same map without the CMB 
(top right), but still including point source and instrument noise.  
The relative importance of these two effects can be seen in the bottom maps, 
where we have displayed the SZE with point sources but no instrument noise 
(left) and vice versa (right).  The maps are 
$3^{\circ} \times 3^{\circ}$ and contain $1024^2$ pixels, rebinned to 
$256^2$ for display.  The color scale of the maps is linear, and 
span $100 \mu$K, except for the map including the primary CMB, 
which spans $500 \mu$K.  This particular map is made at 150 GHz for a 12 
meter dish, assuming $10 \mu$K-arcmin of instrument noise, and with point 
source contributions near the high end of the expected magnitude.}
\label{fig:maps}
\end{figure}
Since we use the method outlined in \cite{SW03} to create maps of the SZE, we 
provide only a brief description here.  The maps are created from a large 
volume, high resolution N-body simulation containing a fair sample of the 
universe, for a flat $\Lambda$CDM cosmology with 
$\Omega_m = 0.3$, $\Omega_b h^2 = 0.02$, $h = 0.7$, and $\sigma_8 = 1$.  
We use a semi-analytic model, in which baryonic matter traces the dark 
matter in our clusters, in order to include the gas physics responsible 
for the SZE.  This assumption is likely to be a good approximation 
everywhere except at the cluster cores, which will not be resolved by the 
surveys considered here.  We identify clusters in the N-body simulation 
using a friends-of-friends (FOF) algorithm \citep{Davis85} with a linking 
length $b = 0.15$ times the mean interparticle spacing.  The mass contained 
by hot gas is set to $\Omega_b / \Omega_m$ of the total, and each cluster 
is set to be isothermal at a temperature given by
\begin{equation} \label{eq:clustertemp}
{k_B T \over keV}  \sim 
\left ( {H(z) M \over 10^{15} h^{-1} M_\odot } \right )^{2 / 3} 
\end{equation}
where $H(z)$ is the hubble parameter.  This effectively reproduces the results 
of the hydrodynamic simulations of \cite{WHS02}.  The normalization has been 
set to pass through the lower envelope of the CBI deep field \citep{Mason03} 
and through the BIMA point \citep{Dawson01} on small angular scales.  We 
generate Compton-Y maps by projecting along each line of sight, so that
\begin{equation} \label{eq:comptony}
y=\int \sigma_T n_{e}{k_B T \over m_{e} c^2} dl \qquad
\end{equation}
where $\sigma_T$ is the Thompson scattering cross section, $n_e$ is the 
electron number density, and $m_e$ is the electron mass.  The temperature 
fluctuation for a given frequency $\nu$ is related to the Y-maps by
\begin{equation} \label{eq:DeltaTonT}
{\Delta T \over T} = y \left ( x {e^x + 1 \over e^x -1} - 4 \right )
\end{equation}
where $x = h \nu / k_B T_{CMB} \simeq \nu / 56.84$GHz is the dimensionless 
frequency.  Ten maps are made in this manner (see Figure \ref{fig:maps} for 
an example), each with $1024^2$ pixels and $3^\circ$ on a side.  These are 
not as accurate as those produced using full hydrodynamic simulations, but 
they allow us to probe a larger volume and therfore provide a better sample 
of large clusters situated in their proper cosmological context.

In the absence of perfect spectral information, confusion due to the primary 
CMB temperature anisotropy and to point sources may impede the detection of 
clusters.  We simulate the former using realizations of a Gaussian 
random fields convolved with the CMB power spectrum computed using 
CMBfast \citep{SeZa96}.  We then add radio and infrared (IR) point sources to 
the maps using the model of \cite{WM04}.  For radio sources, this 
is a fit to the Q-band data of WMAP \citep{Bennett03}, while IR 
sources are fit using the 350 GHz observations of \cite{Borys03} 
with the Submillimeter Common User Bolometer Array 
\citep[SCUBA:][]{Holland99} on the James Clerk Maxwell Telescope.  We 
note that there is substantial uncertainty in extrapolating these fits 
to frequencies relevant to us here, so we examine two different 
extrapolations likely to span the magnitude of the effect.

The maps are then smoothed with a Gaussian beam, and Gaussian white noise is 
added.  Although we have ignored many effects which may be important in real 
world observations (such as offsets, drifts, and atmosphere), we treat the 
maps as completed and ready for signal processing.  

\section{Filtering methods} \label{sec:wavelets}

In this section, we describe the filters we use to process the simulated 
maps discussed in Section \ref{sec:simulations}.  These maps have a 
complex structure which makes it impossible to analytically determine 
the best filter to aid our efforts at cluster identification, so we 
explore several different methods.  We begin by describing 
the optimal Fourier based filter of \cite{Tegmark98}, 
then briefly describe filtering in the discrete wavelet basis, where we 
focus on the Daubechies wavelet family \citep{Daub92}.  Finally, we 
discuss filtering using the continuous mexican hat wavelet filter 
\cite[see][for more discussion of filtering with continuous 
wavelets]{Pierpaoli04}.

\begin{figure}
\begin{center}
{\includegraphics*[height=4.8cm]{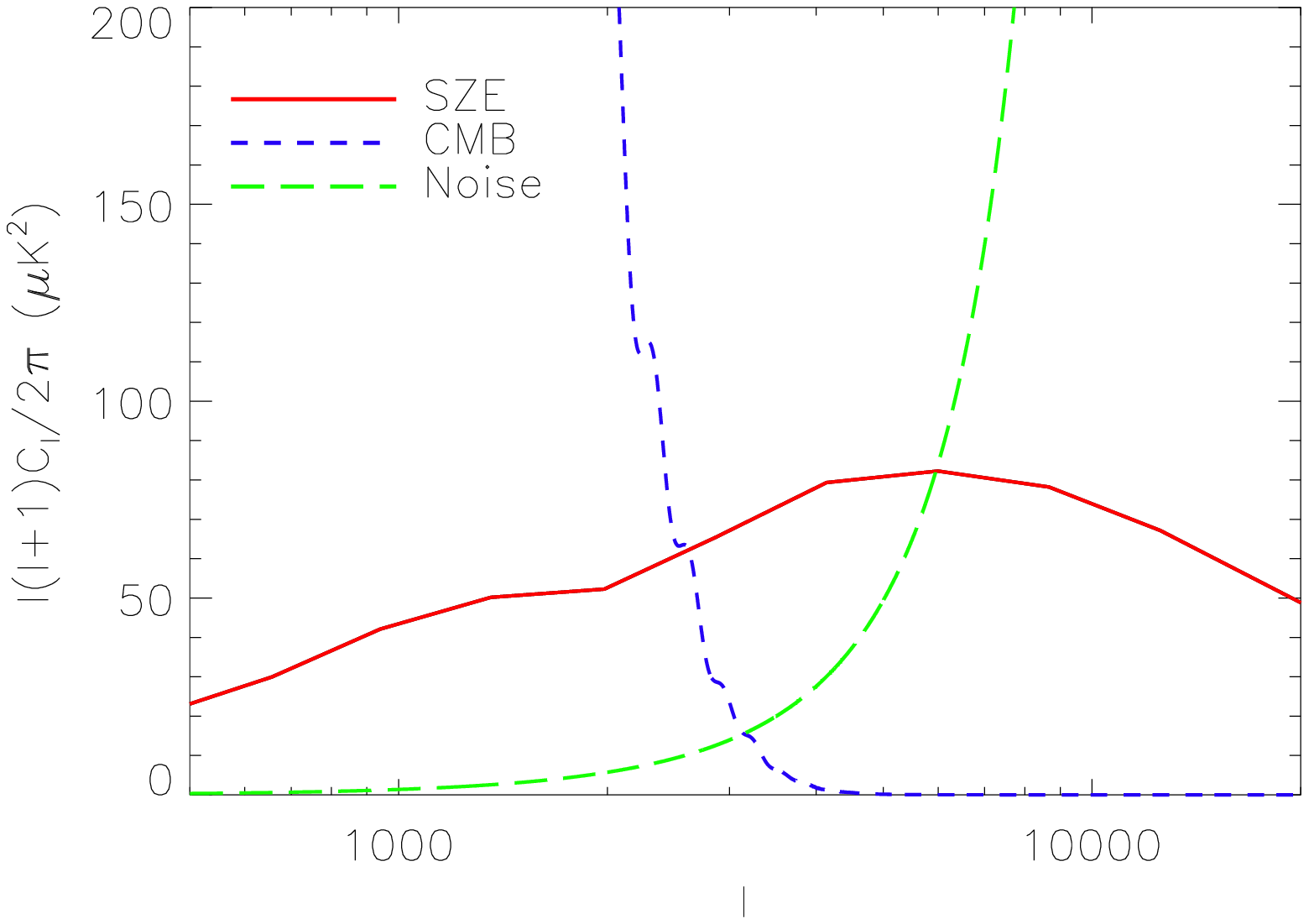}
\includegraphics*[height=4.8cm]{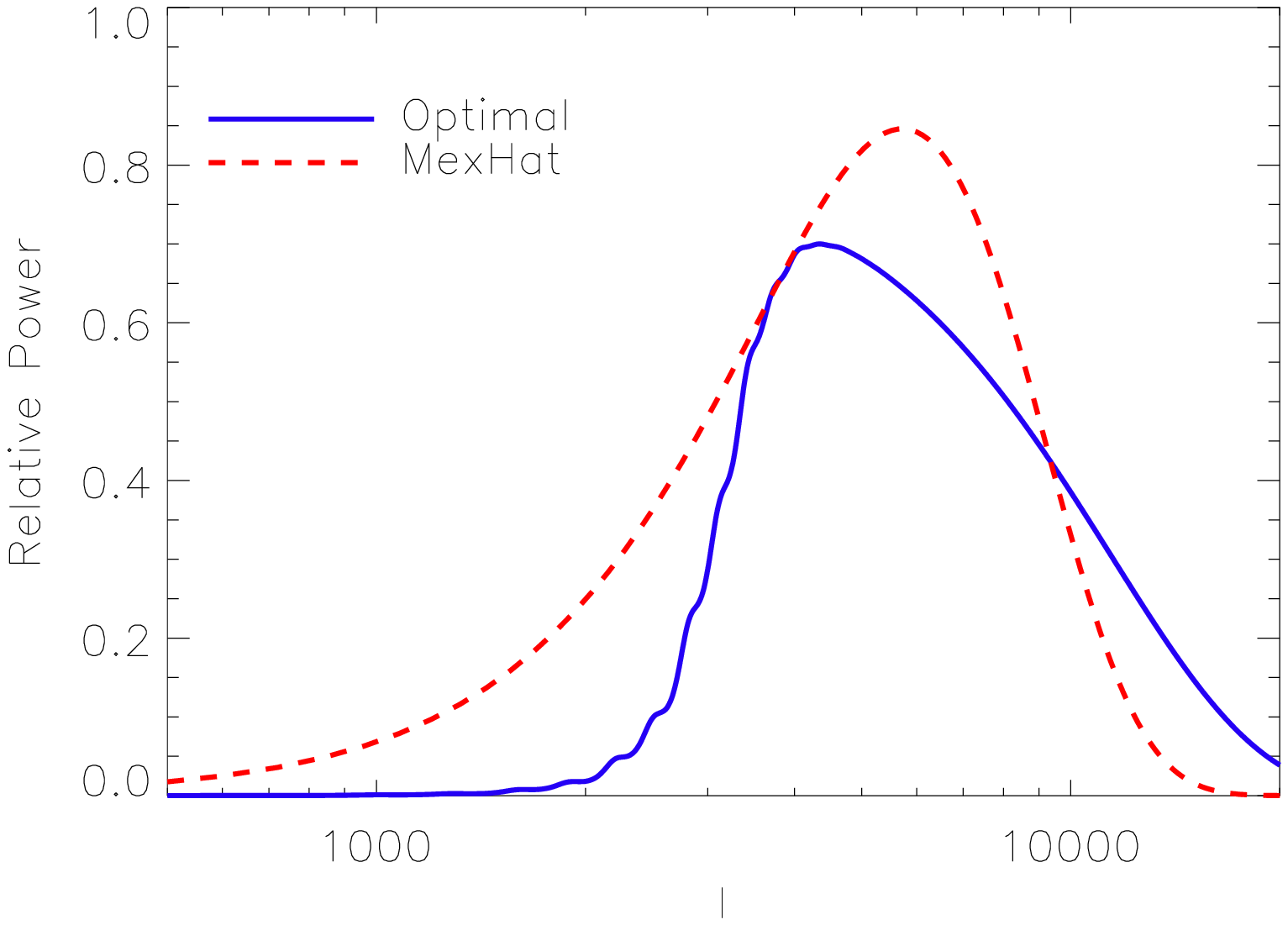}}
\end{center}
\caption{(Left) The expected magnitude of the SZE, the CMB, and 
instrument noise for a fiducial $10 \mu K$ per $1^{\prime}$ beam survey, 
shown here in Fourier space.  The SZE signal exceeds both the noise and 
CMB on roughly arcminute scales.  (Right) The optimal and mexican hat 
filters in Fourier space.  The filters are wedge shaped band pass filters, 
designed to pass scales where the SZE is large relative to the CMB and 
instrument noise.}
\label{fig:szwindow}
\end{figure}

The signal power in the SZE is expected to exceed that of the primary CMB 
and (for a sufficiently powerful survey) instrument noise on roughly arcminute 
scales (Figure \ref{fig:szwindow}).  The optimal filter derived in 
\cite{Tegmark98} 
is essentially a bandpass window which can be centered on the appropriate 
angular scale, and is therefor an obvious candidate for our purpose.  
This filter is azimuthally symmetric, and its radial dependence in 
Fourier space is
\begin{equation} \label{eq:tegfilt}
\tilde{\psi}_{match}(\ell) \sim 
{e^{\theta^2 \ell(\ell +1)/2} \over C_{\ell}^{\rm{Tot}} }
\end{equation}
where $\theta$ is the full width half max (FWHM) beam size and 
$ C_{\ell}^{\rm{Tot}}$ is the total power spectrum of all ``noise''.  
See Figure \ref{fig:szwindow} for an example of this filter appropriate 
for our fiducial surveys.

Wavelets have emerged as a powerful tool for signal processing 
\cite[see the Appendix for a brief discussion and e.g.][for a 
review of wavelet signal processing]{Mallat99}.  
They are simultaneously (but imperfectly) localized in both real space and 
Fourier space, and are therefore a natural choice for processing data 
which possesses both real space and Fourier space correlations, such 
as we expect from our fiducial surveys.  We employ the wavelet transform 
algorithm outlined in \cite{Press92}, and although we focus on the 
Daubechies wavelets, we note that we have also explored the Coiflet, 
Symlet, and Morlet wavelet families, and as these offer essentially 
the same results as the Daubechies wavelets, we do not discuss 
them further.  

To generate our filter, we follow \cite{Pen99} and estimate the 
expected signal given data $\langle S | D \rangle$ for each coefficient in 
the wavelet transform.  This amounts to Wiener filtering 
in the limit of Gaussian noise and Gaussian signal, and to 
thresholding for highly non-Gaussian signal distributions, such as unsmoothed 
point sources.  This approach is likely to be superior (in the sense of 
minimizing the least squared error in the reconstruction) if the 
signal or noise include substantial non-Gaussian behavior, but requires that 
the signal and noise probability density functions (PDF) be known.  
Although \cite{Pen99} suggests computing $\langle S | D \rangle$ directly 
from the non-Gaussian behavior of the observed maps, we find that the 
procedure outlined there is subject to numerical artifacts due to the 
finite size of sky in our simulations, so we compute this function directly 
from our 10 input signal maps.  This will not be possible for real surveys, 
and so should be considered an upper limit to the performance of the 
technique.  

We implement a third class of filter, the continuous wavelet filter, using 
the mexican hat wavelet transform \citep[e.g. ][]{Cayon00,Maisinger04}.  The 
continuous wavelet transform $W(a,b)$ of a one dimensional function $f(x)$ 
is a real space convolution of $f(x)$ with a ``mother wavelet'' $\psi(x)$
\begin{equation} \label{eq:continuous}
W(a,b) = \int dx f(x) {1 \over \sqrt{a}} \psi \left ( {x - b \over a} \right )
\end{equation}
where $a$ and $b$ are the scale and position parameters.  This convolution 
is normally performed as a multiplication in the Fourier domain, so that 
$\tilde{W} = \sqrt{a} \tilde{f}(\ell) \tilde{\psi}(a \ell)$, 
where the tilde denotes the Fourier transform.  The mexican hat wavelet is 
the second derivative of a Gaussian, so that its Fourier transform is
\begin{equation} \label{eq:mexhat}
\tilde{\psi}_{mex}(a \ell) \sim ( a \ell)^2 \rm{exp} 
\left [ - {(a \ell)^2 \over 2} \right ]
\end{equation}
Like the optimal filter, the continuous wavelet transform is essentially 
a filter in Fourier space, as can be seen in the side by side comparison 
in Figure \ref{fig:szwindow}.  

In the next section, we demonstrate the use of these filters on the simulated 
maps described in Section \ref{sec:simulations} for various survey 
strategies.
\begin{figure}
\begin{center}
{\includegraphics*[height=6.8cm,width=6.8cm]{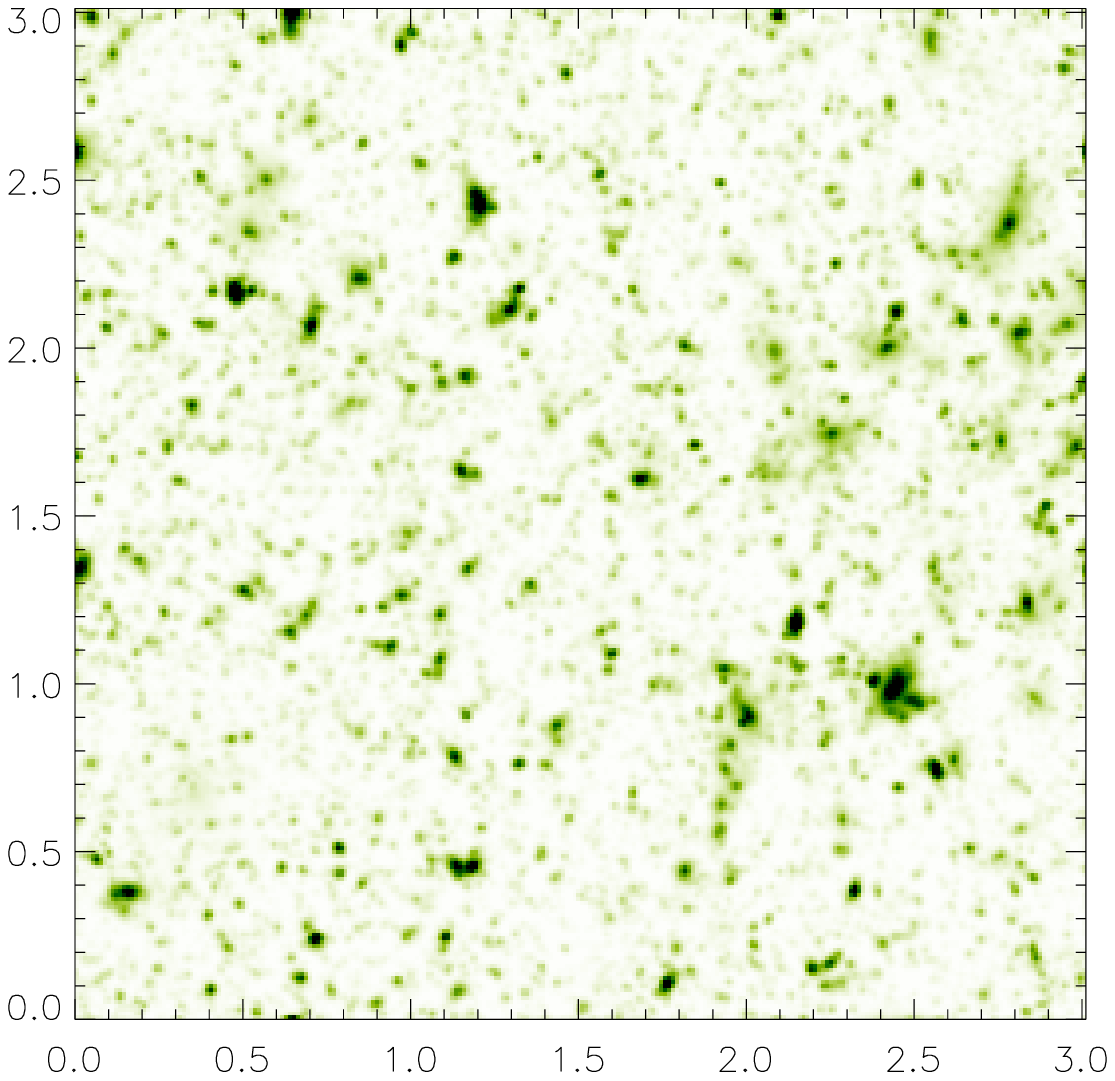}
\includegraphics*[height=6.8cm,width=6.8cm]{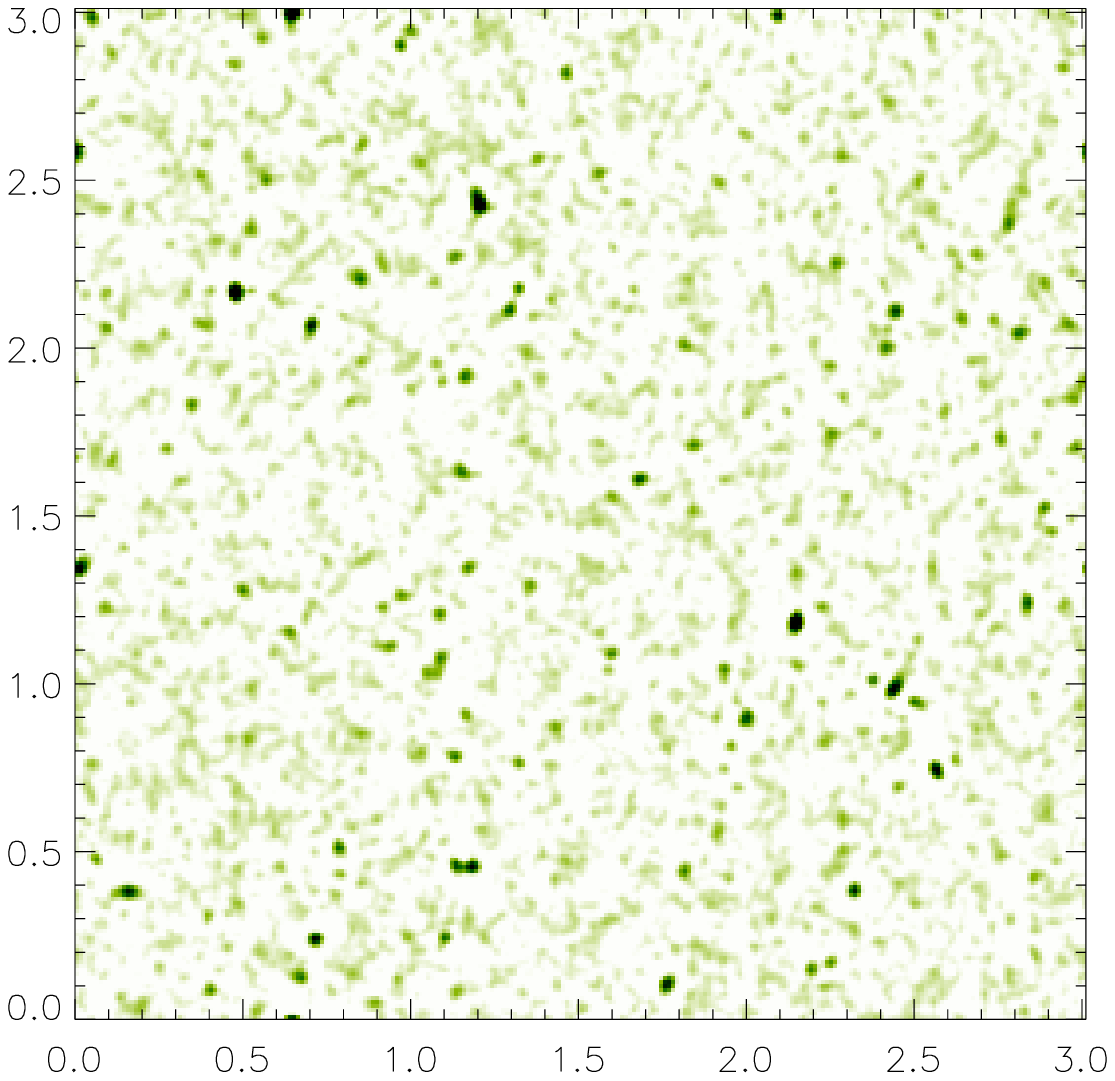}}
{\includegraphics*[height=6.8cm,width=6.8cm]{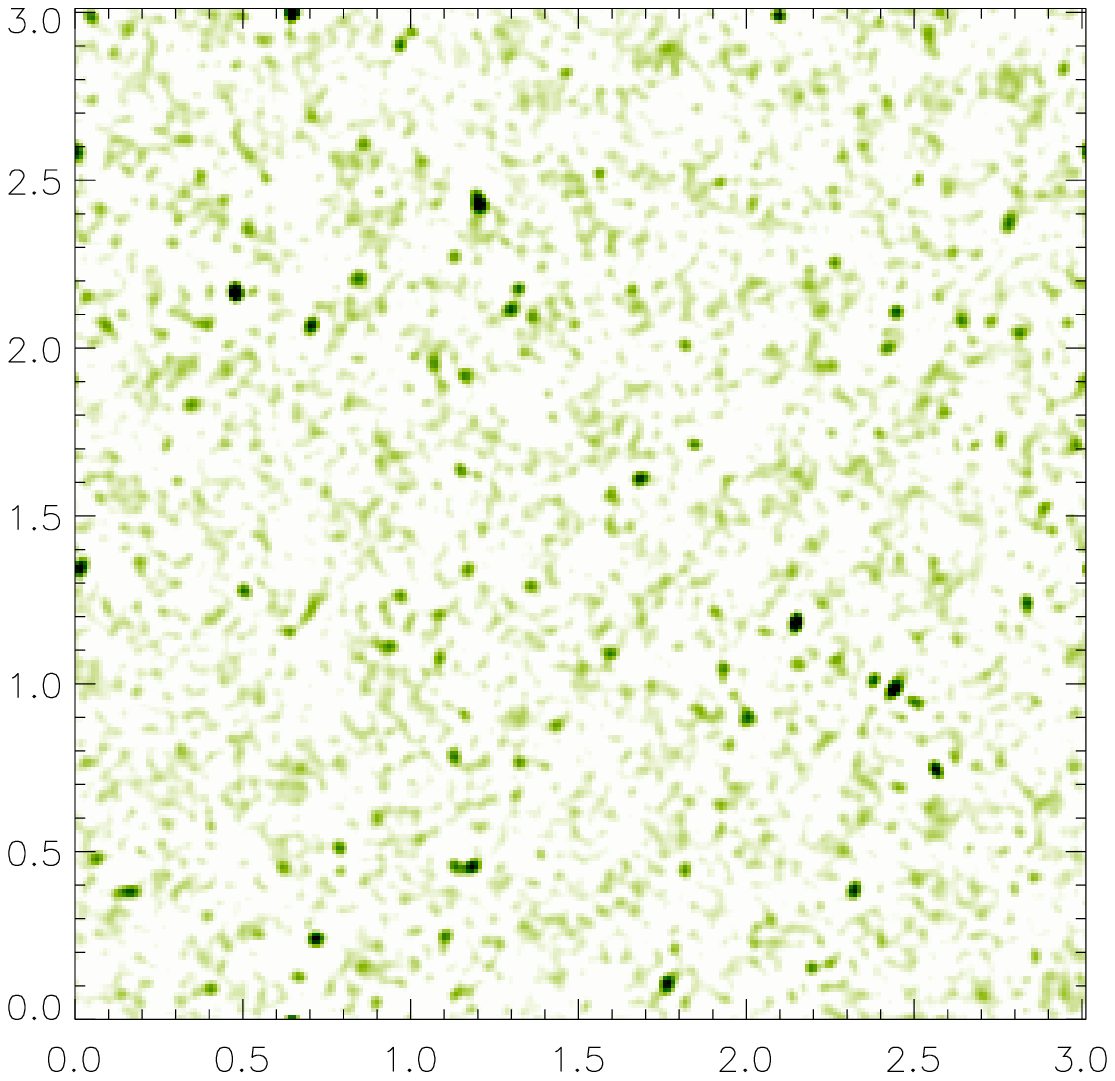}
\includegraphics*[height=6.8cm,width=6.8cm]{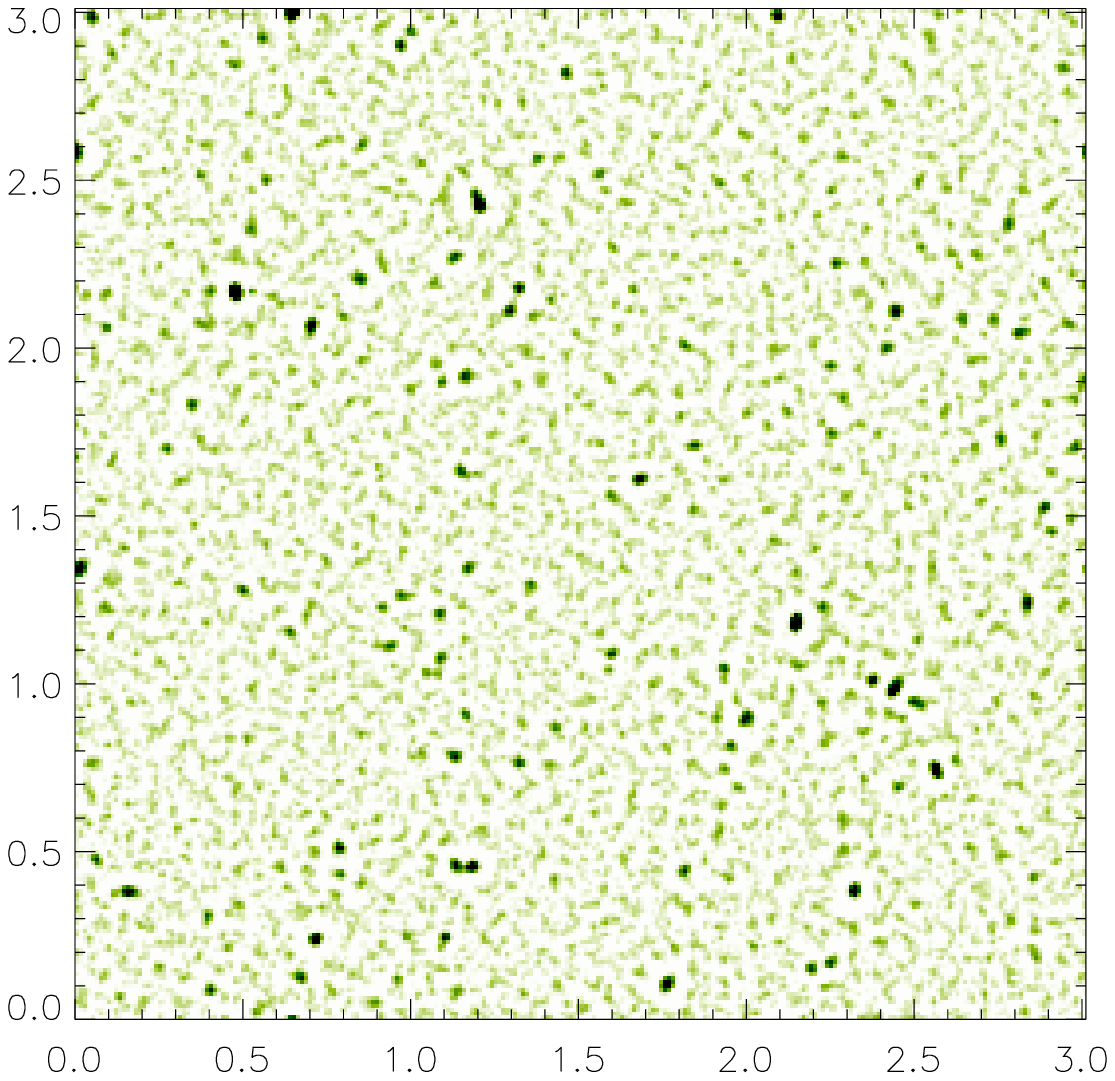}}
\end{center}
\caption{The input SZE (top left) and the filtered maps for the discrete 
Daubechies wavelet filter (top right), the mexican hat continuous wavelet 
filter (bottom left), and for the optimal Fourier filter (bottom right).  
The color scale is linear, and structures 
less than $1 \sigma$ of the noise have been suppressed for visual clarity.}
\label{fig:filtermaps}
\end{figure}

\section{Results} \label{sec:results}

In this section, we examine the maps after application of the filters 
described in Section \ref{sec:wavelets}.  As can be seen in the examples 
shown in Figure \ref{fig:filtermaps}, 
the three filters clearly all succeed in improving signal 
to noise, with large structures in the filtered maps all corresponding to 
massive clusters.  We quantify the level of this success using the peak 
finding algorithm of \cite{SW03}, and conclude that for the surveys we 
consider here, the optimal filter performs at least as well as the 
wavelet based filters for creating complete, efficient surveys of clusters.  
We then examine cluster finding in the context of several survey strategies.

Of the three filtered maps shown here, the one created using the discrete 
wavelet (specifically, Daub6) ``$\langle S | D \rangle$'' technique 
described in Section \ref{sec:wavelets} best reconstructs the input SZE 
signal in the sense of minimizing the least squared error of the 
reconstruction.  However, this reconstruction does not do as well at 
creating complete, efficient surveys of clusters.  This is because a 
large fraction of the improved signal recovery is associated with a few 
giant clusters.  Since these produce enormous signal, they are easy to 
find, regardless of the filter, and improving their reconstruction 
does not aid in the completeness of the catalogue.  However, these 
clusters have signal on relatively large angular scales, so that the 
``$\langle S | D \rangle$'' technique smooths the maps more than is 
optimal for finding smaller clusters at the threshold of detection.  
Although this can be accounted for by eliminating the large clusters from 
consideration when formulating the filter, the result is to simply 
scale the signal by a constant at each level, so that the smaller 
cluster signal is effectively in the Wiener filter limit.  Although we do 
not show them here, we have also considered maps with holes, rough edges, 
and spatially varying noise.  The wavelet filter performs better in this 
context than the other filters for extreme conditions, but there is no 
detectable advantage for realistic assumptions.  
\begin{figure}
\begin{center}
{\includegraphics*[height=5.8cm]{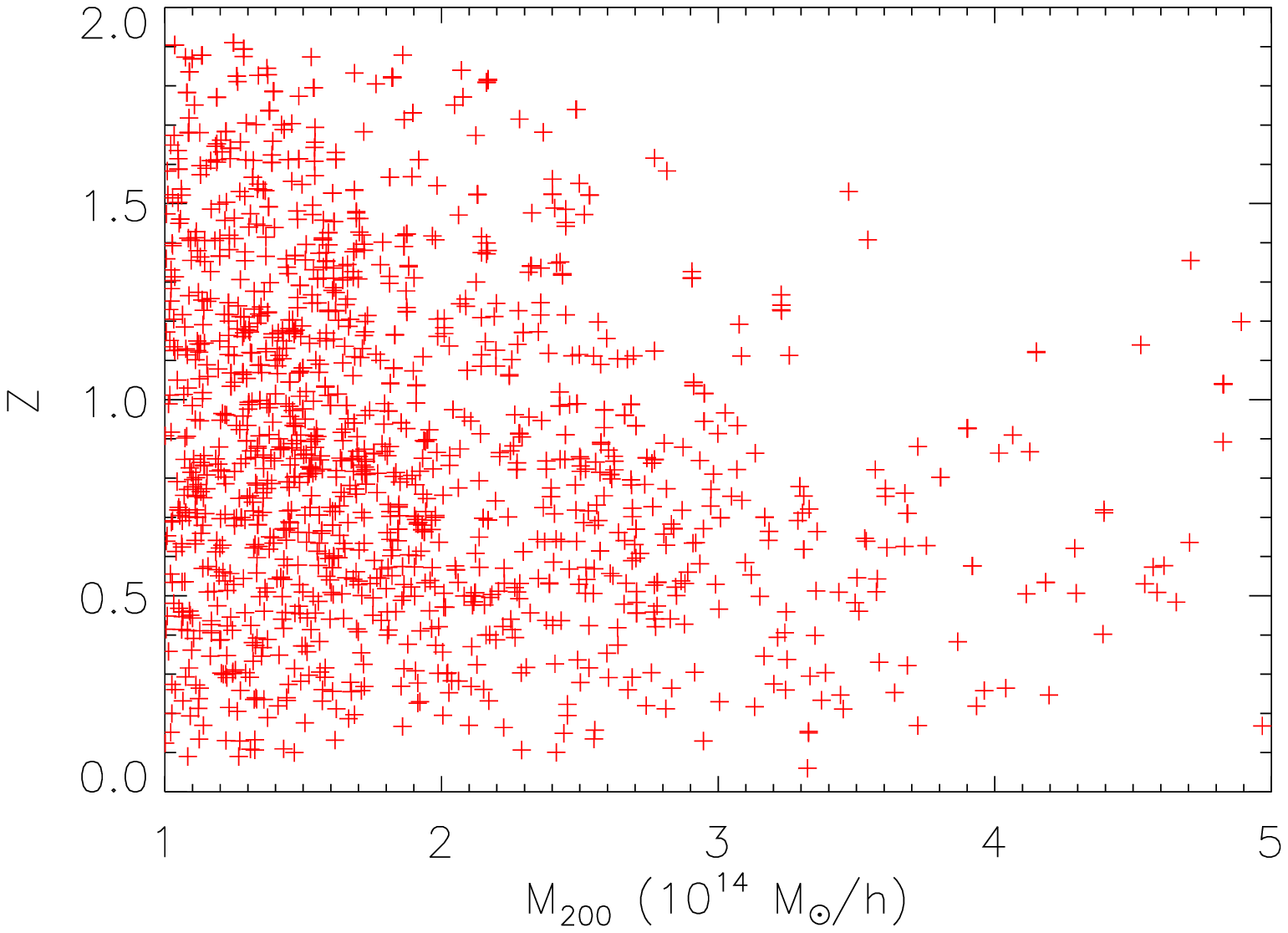}
\includegraphics*[height=5.8cm]{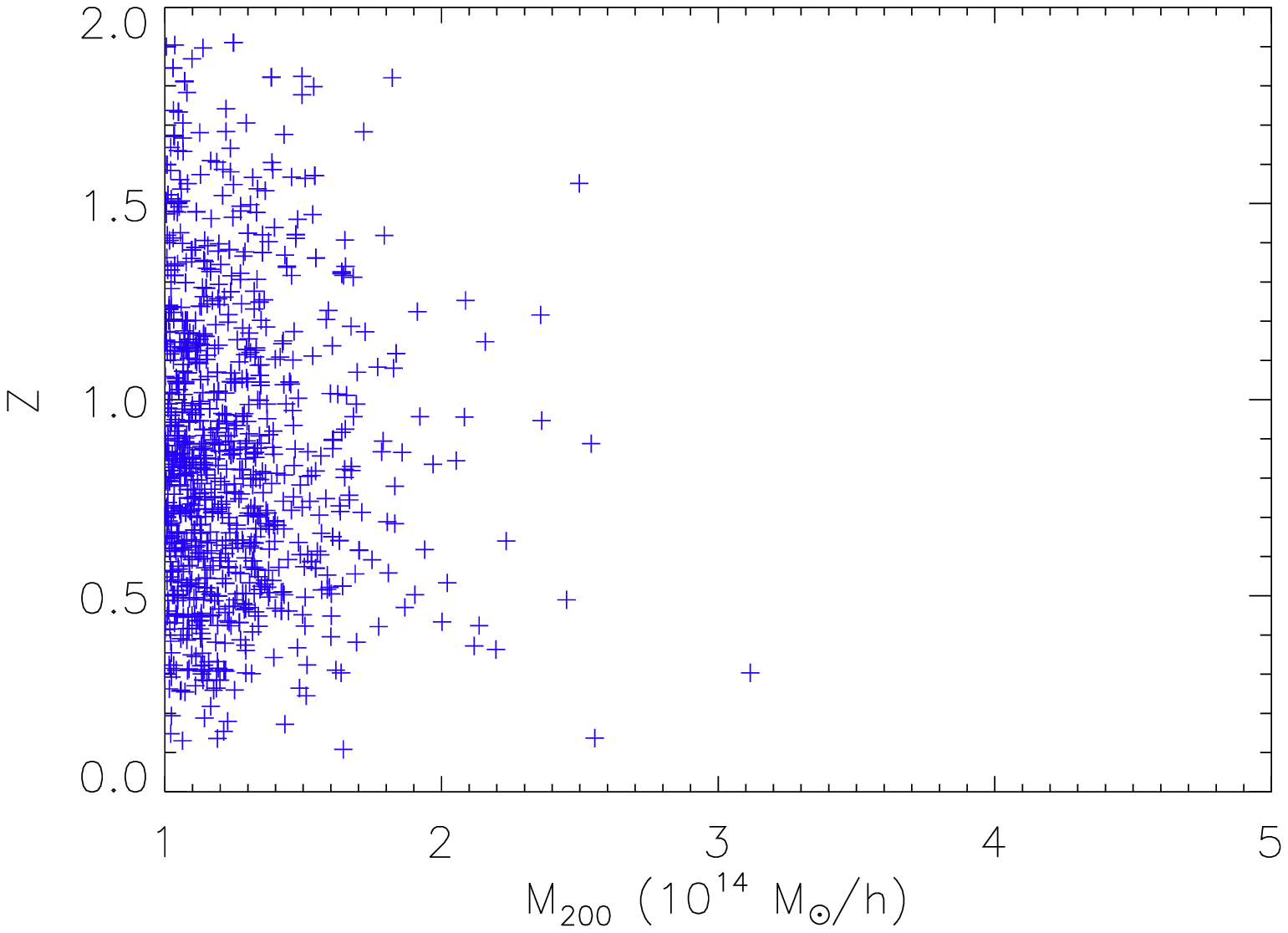}}
\end{center}
\caption{Clusters found (left) and missed (right) as a function of mass and 
redshift for our fiducial $10 \ \mu$K per arcminute beam survey.  The survey 
size is 90 square degrees, and we have required a 75\% detection efficiency 
for clusters of mass above $10^{14} \ \rm{M}_\odot$.}
\label{fig:foundmissed}
\end{figure}

\begin{figure}
\begin{center}
{\includegraphics*[height=8.5cm,width=5.5in]{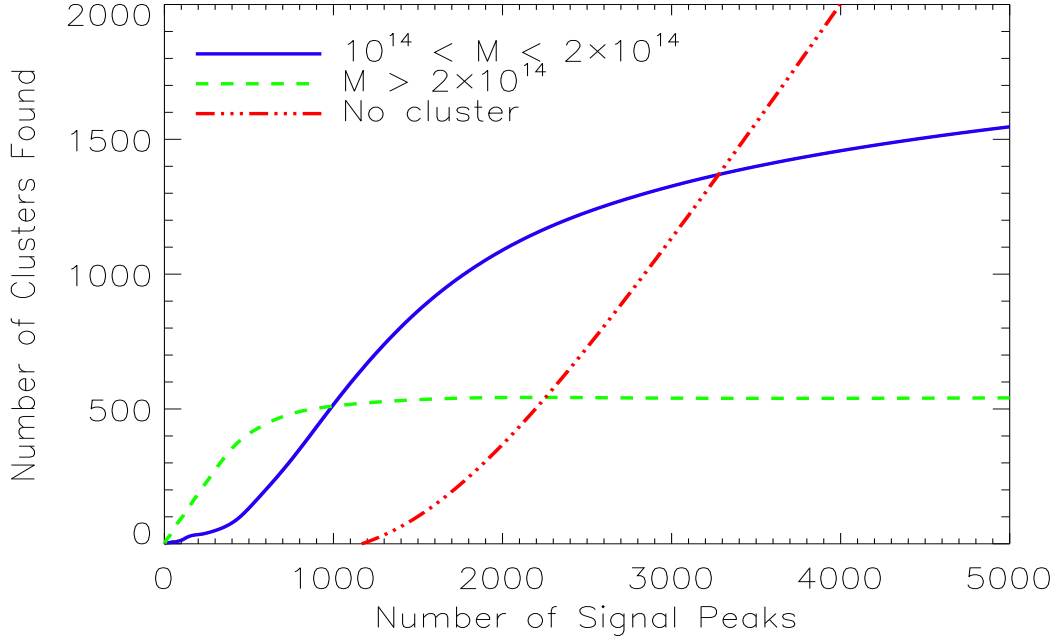}}
\end{center}
\caption{The number of massive clusters associated with the 5000 largest 
peaks in the filtered signal for our 10 maps, shown here in descending 
order of signal strength, for our fiducial $10 \ \mu$K per arcminute 
beam survey.  The largest signal peaks are always associated 
with a cluster.  However, noise and projection effects cause 
substantial scatter in the mass-observable relation.  In the small 
signal regime, this scatter begins to dominate, so that many 
smaller peaks are not associated with a cluster of mass 
$\rm{M} > 10^{14} \ \rm{M}_\odot$.  We note that there are a total of 
1900 clusters in the (solid) mass range of 
$10^{14} < \rm{M} < 2 \times 10^{14} \ \rm{M}_\odot$ 
and 550 of (dashed) mass $ \rm{M} > 2 \times 10^{14} \ \rm{M}_\odot$ 
in our simulation.}
\label{fig:completeness1}
\end{figure}

\begin{figure}
\begin{center}
{\includegraphics*[height=8.5cm,width=5.5in]{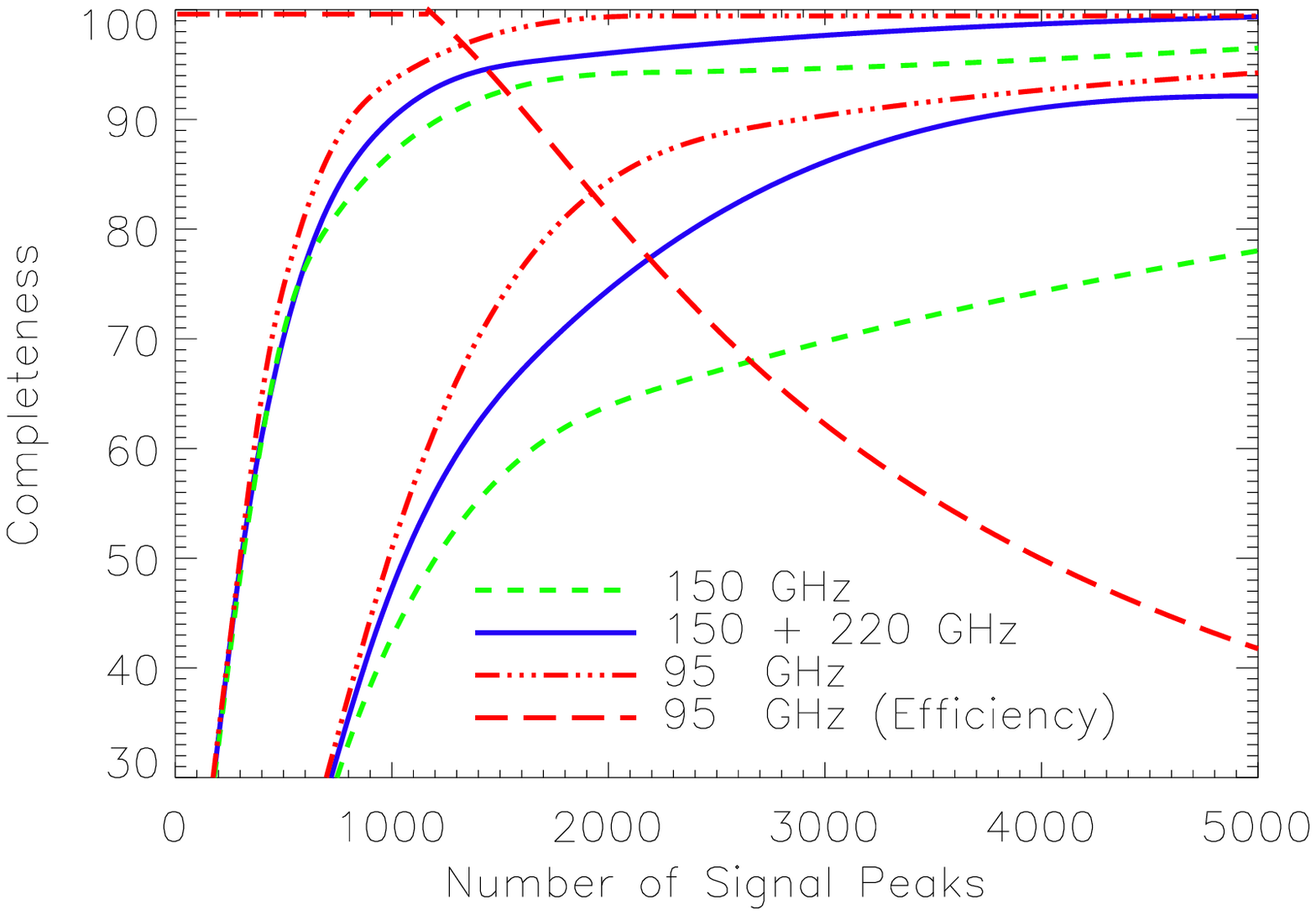}}
\end{center}
\caption{Completeness of clusters detected above a mass threshold for three 
survey strategies as a function of the number of signal peaks included in the 
analysis.  The upper family 
of curves is the completeness for clusters of mass greater than 
$2.0 \times 10^{14} \ \rm{M}_\odot$, while the lower family is for clusters of 
mass less than this but greater than 
$1.5 \times 10^{14} \ \rm{M}_\odot$.  The efficiency for a threshold mass of 
$10^{14} \ \rm{M}_\odot$ for one survey is also shown.  The totals are derived 
from ten $3^{\circ} \times 3^{\circ}$ simulated maps.}
\label{fig:completeness}
\end{figure}
The mexican hat filter and the optimal filter of \cite{Tegmark98} perform 
about equally well.  In fact, we find that the primary CMB anisotropy is 
easy to separate morphologically from the SZE for the small beam sizes 
we consider here, and any reasonable hi-pass filter can be used for 
this task.  If the two filters are then set to be roughly the same on small 
angular scales, then they perform about equally well at cluster finding.  In 
particular, we find that for clusters which are smaller than 
the beam of the survey, simply smoothing the noisy maps by the beam 
performs best for small scale filtering.

We use two methods to identify signal peaks in the filtered maps.  The 
first simply flags local maxima, while the second computes the total 
flux in all pixels surrounding (and including) the local maxima which are 
greater than one quarter the peak value.  Since the choice of method does 
not significantly alter our results, the results presented here are for the 
simpler local maxima technique unless stated otherwise.

Once we have identified signal peaks in the maps, we compare these to a 
list of clusters in our simulation.  As can be seen in 
Figure \ref{fig:foundmissed}, massive clusters are nearly always 
identified using this method, while no particular dependence on 
redshift is evident.  Also, large signal peaks are nearly always 
associated with a cluster (Figure \ref{fig:completeness1}), although 
noise and projection effects cause substantial scatter in the 
mass-observable relation.  The efficiency of the survey (that is, 
the chance that a signal peak corresponds to a cluster) is therefor 
nearly 100\% for large peaks.  

We now turn our attention to survey strategy, where we examine the 
use of multiple vs single frequency measurements.  Although the primary CMB 
anisotropy is not a serious contaminant, contributions from point sources 
are non-negligible.  Spectral information can alleviate this issue, but at 
the price of either reduced signal to instrument noise, or less sky coverage, 
per unit of telescope time.  If only a single frequency is used, then point 
source contributions must be considered in addition to signal strength, 
instrument noise, and beam size.  

To investigate this issue, we begin with a fiducial survey at 150 GHz for a 
12 meter dish (roughly a $0.^{'} 8$ beam at 150 GHz) and $10 \mu$K-arcmin 
instrument noise over our 90 square degrees of simulated sky.  As expected, 
this survey does well at cluster identification in the absence of point 
sources.  However, including them substantially worsens the result 
(Figure \ref{fig:completeness}), even for a level at the low end of the 
expected confusion noise, and a deeper integration at the same frequency only 
marginally improves the outcome.  A better result is achieved by combining 
the 150 GHz survey with a 220 GHz observation and differencing the results to 
remove the point source contribution.  Although the 220 GHz channel contains 
no signal and is noisier than the 150 GHz channel 
(we assume $\sim 15 \mu$K-arcmin for the same integration time), the 
effective signal to instrument noise is only marginally worse.  This is 
because the point source contribution is roughly twice as large at 220 GHz as 
it is at 150 GHz, so that (roughly speaking) the point source elimination is 
accomplished by including only half the noise of the instrument noise at 220 
GHz.  Added in quadrature, this is effectively a $12 \mu$K-arcmin survey.
\begin{figure}
\begin{center}
{\includegraphics*[height=5.3cm]{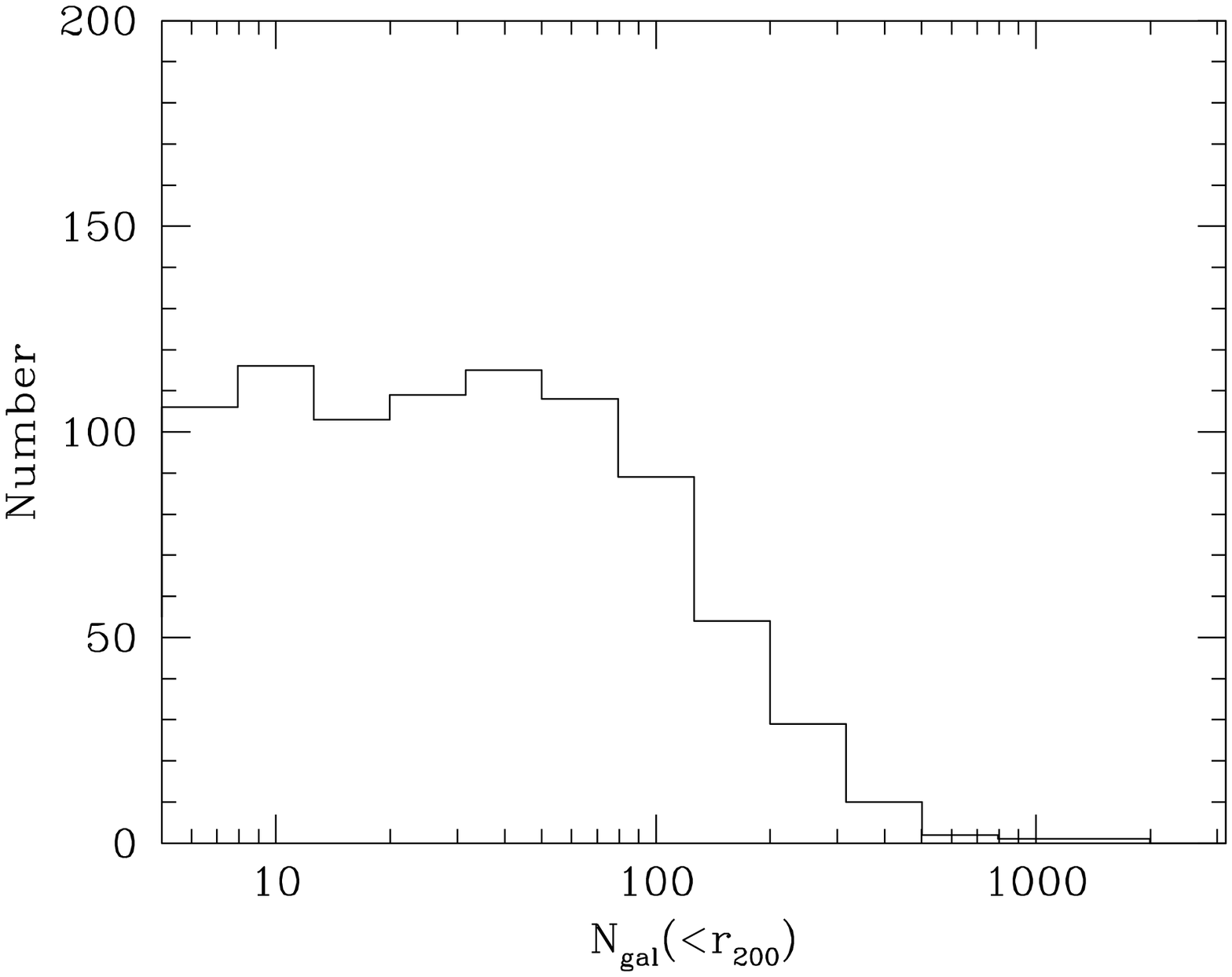}
\includegraphics*[height=5.3cm]{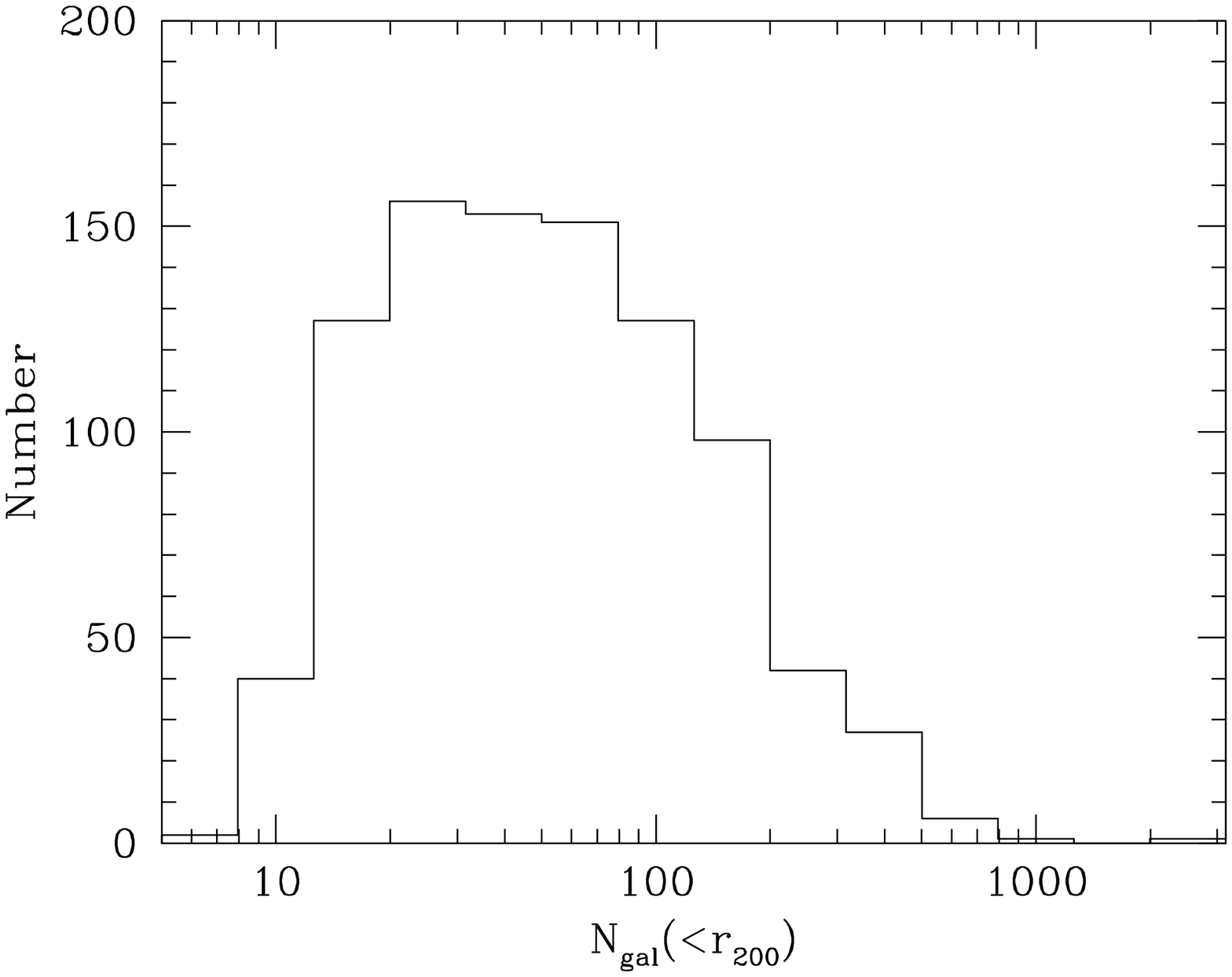}}
\end{center}
\caption{The number of galaxies brighter than (left) R = 24 and 
(right) R = 25 within the virial radius of clusters in our sample 
with mass greater than $10^{14} \rm{M}_\odot$ and at redshift 
$Z<1$.  The number within the core region, where the contrast 
against the background is highest, is about 10\% of $N_{gal}(< r_{200})$.}
\label{fig:ngal24}
\end{figure}

An alternative strategy is explored in the form of a single frequency 
survey at 95 GHz.  This offers the advantage of a larger signal and lower 
point source contamination, but is noisier (we assume $17 \mu$K-arcmin for 
equal integration time) and has a larger ($\sim 1.^{`} 3$) beam.  This 
results in improved detections, and although the level of improvement 
depends upon the model of point source confusion assumed, this survey 
strategy is superior in our simulations for the models we have considered 
here.  We have not accounted for clustering of point sources, 
which may change this result.

Finally, we note that optical and near-IR emission is still the least 
expensive way of measuring cluster redshifts, and that redshift 
information is crucial to the physical interpretation of the cluster 
sample.  We use the method outlined in \cite{White03} to provide a 
rough estimate (Figure \ref{fig:ngal24}) of the number of galaxies 
brighter than $R=24$ and $R=25$ within the virial radius of the 
clusters detected (at 75\% efficiency above mass 
$M > 10^{14} \ M_{\odot}$ and at redshift $Z < 1$) in our fiducial 
survey.  The number in the core region, where the contrast against the 
background is the highest, will obviously be smaller.  If the galaxies 
follow the mass, approximately 20\% of the galaxies lie within the break 
radius ($0.2r_{200}$) and 8\% within the core radius ($0.1r_{200}$).  
Although the results we show here are for clusters at redshift $Z < 1$, 
the results for higher redshifts will clearly be somewhat worse.  
Optical follow up will no doubt be an integral but challenging complement 
to an SZE cluster survey similar to that discussed here.

\section{Conclusions} \label{sec:conclusions}

Measurement of the CMB using the unprecedented combination of power, 
resolution, and sky coverage expected in upcoming surveys will return a 
wealth of information, including high resolution detections in the SZE sky 
sufficient to provide enormous catalogues of galaxy clusters.  We have 
studied the use of several filtering techniques to aid the cluster 
identification process, and evaluated the likely detection of clusters 
using simulated maps of the CMB.  We note that while further signal analysis 
will be required to optimally measure cluster properties, we have not 
addressed that issue here.

We have tested three filter techniques, using Fourier methods, continuous 
wavelets, and discrete wavelets, and have found that all of these can 
be efficiently used to enhance the signal to noise in our maps.  Although 
the discrete wavelets perform better under some extreme conditions, we 
find that each of these techniques may be used effectively to aid cluster 
detection.  

We have also examined the success of survey strategies in creating complete 
catalogues of clusters for a given mass threshold.  We find that the primary 
CMB anisotropy is not an important source of noise for the high resolution 
surveys we have considered here, but that point sources may in some cases be 
more important than instrument noise.  Accordingly, multi-frequency 
measurements are likely required if the sky is probed in frequency bands 
where point sources are large relative to the signal strength.  A single 
frequency band may be used effectively if the band center is selected at a 
frequency where point source contamination is not expected to overwhelm the 
signal.  We note that the magnitude of the point source confusion and 
clustering are not as yet well measured at frequencies relevant to us here, 
so that we are dependent upon models for our results.  

A well understood, nearly complete catalogue of massive clusters over a 
large fraction of the sky would be a major achievement for cosmology, and 
will likely be available in the near future as powerful surveys begin 
operation.  The road to optimizing the results includes a determination 
of the best survey strategies and signal processing techniques, and in this 
endeavor, simulations can play an important role.  We have made the raw 
maps, along with some auxiliary data products, freely available to the 
community at http://mwhite.berkeley.edu/ in the hope that they will be 
useful in taking the next step.

CV would like to thank J.D. Cohn, Tom Crawford, Steve Myers, and Wayne Hu 
for useful discussions.  The simulations used here were performed on the 
IBM-SP at the National Energy Research Scientific Computing Center.
This research was supported by the NSF and NASA.

\appendix \section{Appendix:  Wavelets} \label{sec:appendix}

In this section, we provide a brief discussion of discrete wavelets to 
orient the reader \citep[see e.g.][for a more substantial introduction to 
wavelet signal processing]{Mallat99}.  
To get a feel for wavelets, let us consider the first order Daubechies 
wavelet, Daub1, also called the Haar wavelet.  In one dimension, the 
first level (not to be confused with order) Daub1 transform involves 
computing the two pixel average 
$\mathbf{a}$ and difference $\mathbf{d}$ of a signal 
$\mathbf{f}  = (f_1, f_2, \dots, f_N)$, so that the elements $a_m$ of 
$\mathbf{a}$ are defined by
\begin{equation} \label{eq:trend}
a_m = {f_{2m-1} + f_{2m} \over \sqrt{2}}
\end{equation}
and similarly for $\mathbf{d}$, but with a minus sign on the right hand side 
of Eq.(\ref{eq:trend}).  Like all discrete wavelet transforms, the Daub1 
transform decomposes a signal into two subsignals half the length of the 
original:  a running average $\mathbf{a}$ called the trend, and a running 
difference $\mathbf{d}$ called the fluctuation.  Note that this transform is 
linear, invertible, and preserves the total sum of squares of the pixels 
(the latter is often called ``conservation of energy'' in wavelet parlance).  
For higher order transforms, the trend and fluctuation subsignals are no 
longer simple averages and differences (for example, the Daub2 transform uses 
a four pixel linear fit rather than a two pixel average), but the basic idea 
is the same.

An essential component of wavelet based analysis is the simultaneous 
processing of data at multiple scales.  This ``Multi-Resolution Analysis'' 
(MRA) is implemented by a hierarchical application of the wavelet transform 
on the data, so that the first level transform, which probes the smallest 
physical (and highest frequency) scales, is applied to the original signal.  
The second level transform is then computed by taking the wavelet transform of 
the first level trend signal, and so on, so that for an $\rm{n}^{th}$ level 
transform, the result is a single trend $\mathbf{a^n}$ and n fluctuations 
$\mathbf{d}^1, \mathbf{d}^2, \dots, \mathbf{d}^n$.  
\begin{figure}
\begin{center}
{\includegraphics*[height=8.5cm,width=5.5in]{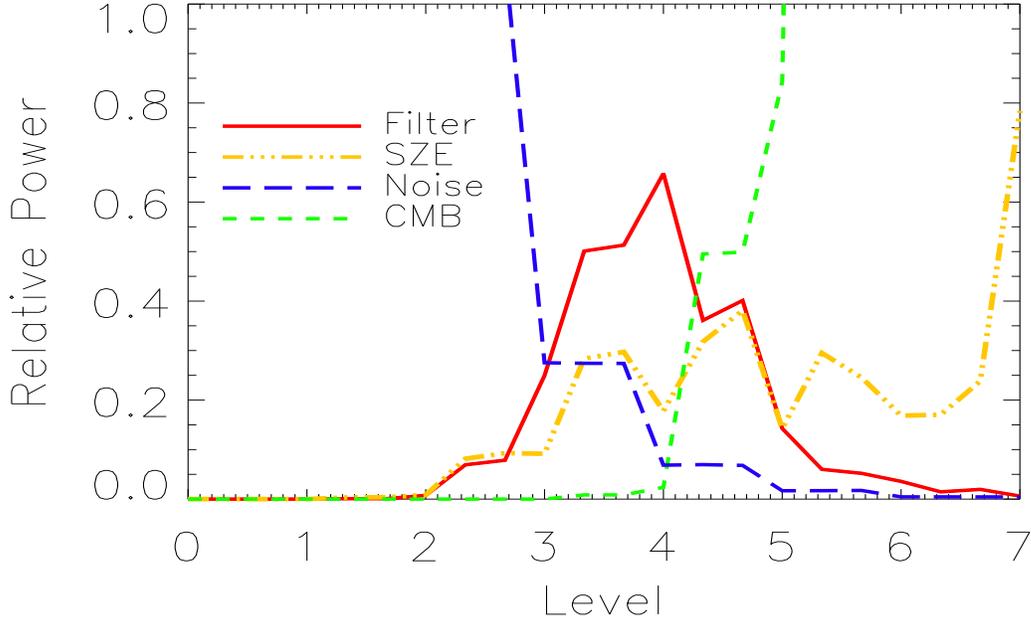}}
\end{center}
\caption{The SZE signal is comparable to the level of confusion from the 
primary CMB and instrument noise on intermediate angular scales.  Here, 
we display this effect for the $6^{th}$ order Daubechies wavelet, and the 
resulting level dependent filter.  }
\label{fig:daub6}
\end{figure}

Most discrete wavelet filtering techniques are based on the ``thresholding'' 
modality, where wavelet coefficients with an absolute value less than a 
chosen threshold value are discarded.  This is effective when the signal is 
much larger than the noise, and when the goal is to recover an image which 
is visually appealing to the human eye.  An approach more suited to our 
purpose is to attempt to reconstruct the signal with a minimum least 
squared error.  For a Gaussian signal and uncorrelated Gaussian noise, 
this implies a Wiener filter in Fourier space.  The equivalent filter in 
wavelet space can be constructed by estimating the energy of the signal 
divided by the data for each level and scaling the transformed data at each 
level by this ratio.  The required estimate of the signal energy can be 
obtained either using simulations, or directly from the data if the 
noise is well understood.  We show an example of this filter for the Daub6 
wavelet in Figure \ref{fig:daub6}.

\begin{figure}
\begin{center}
{\includegraphics*[height=4.5cm,width=4.5cm]{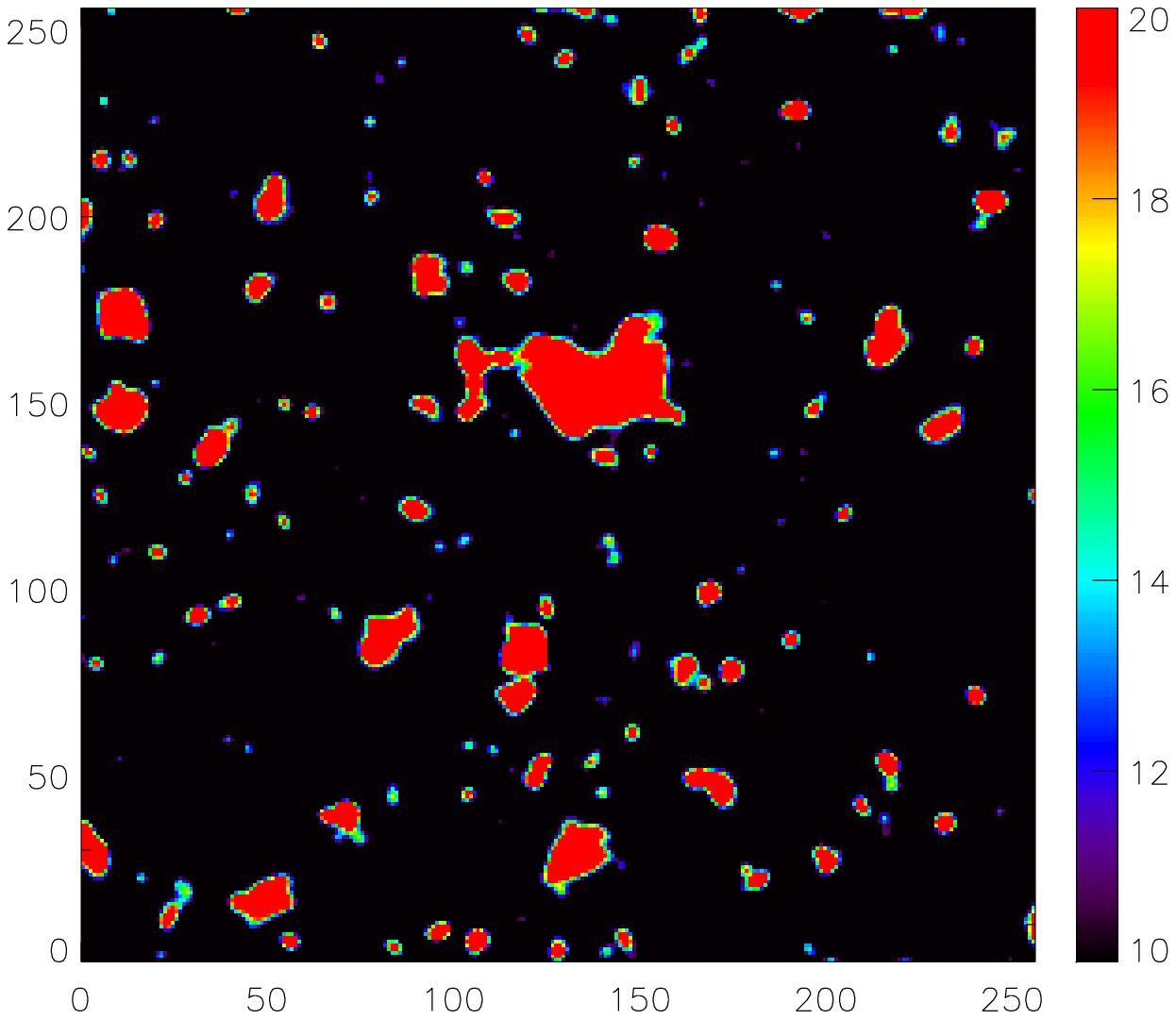}
\includegraphics*[height=4.5cm,width=4.5cm]{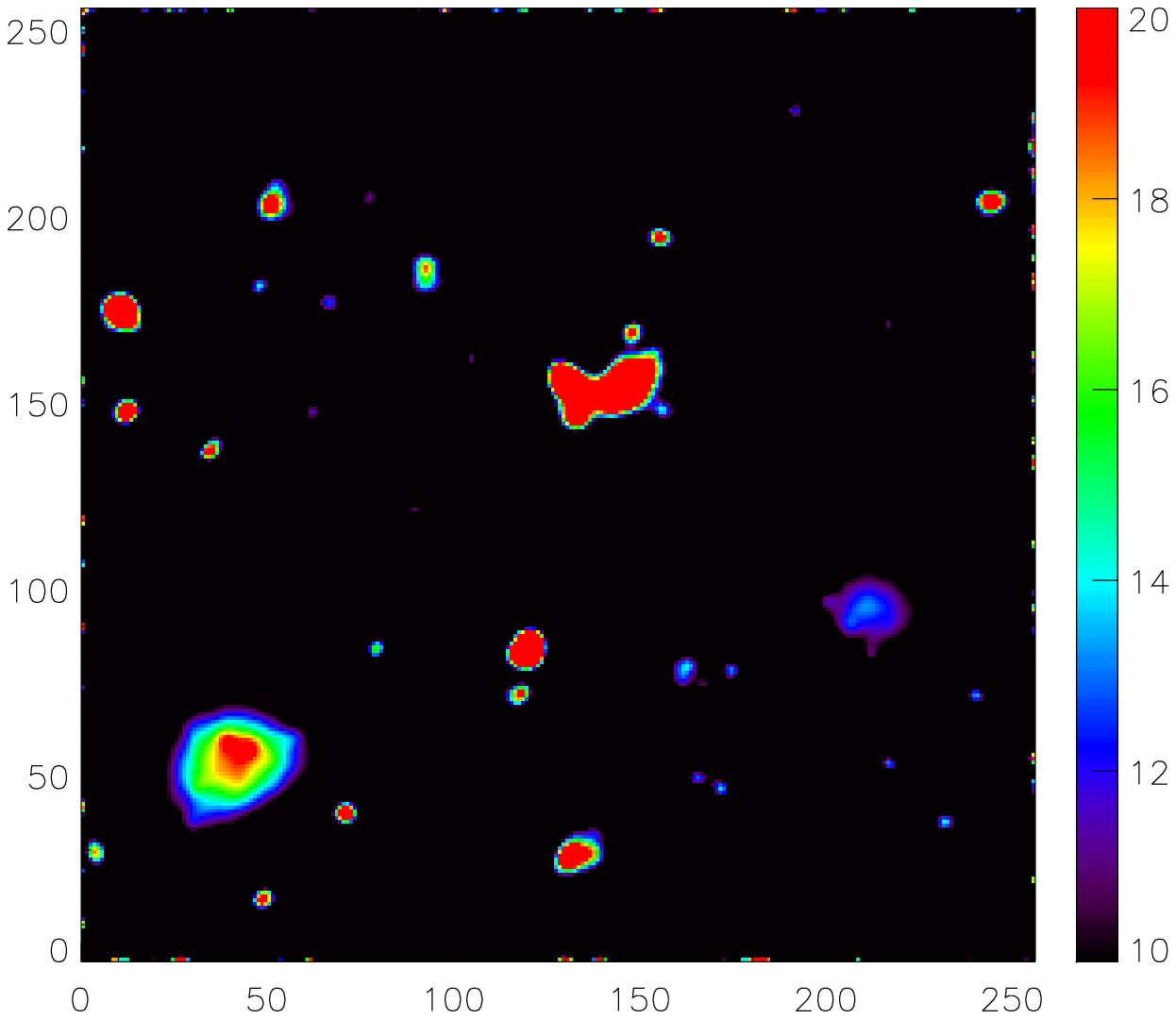}}
{\includegraphics*[height=4.5cm,width=4.5cm]{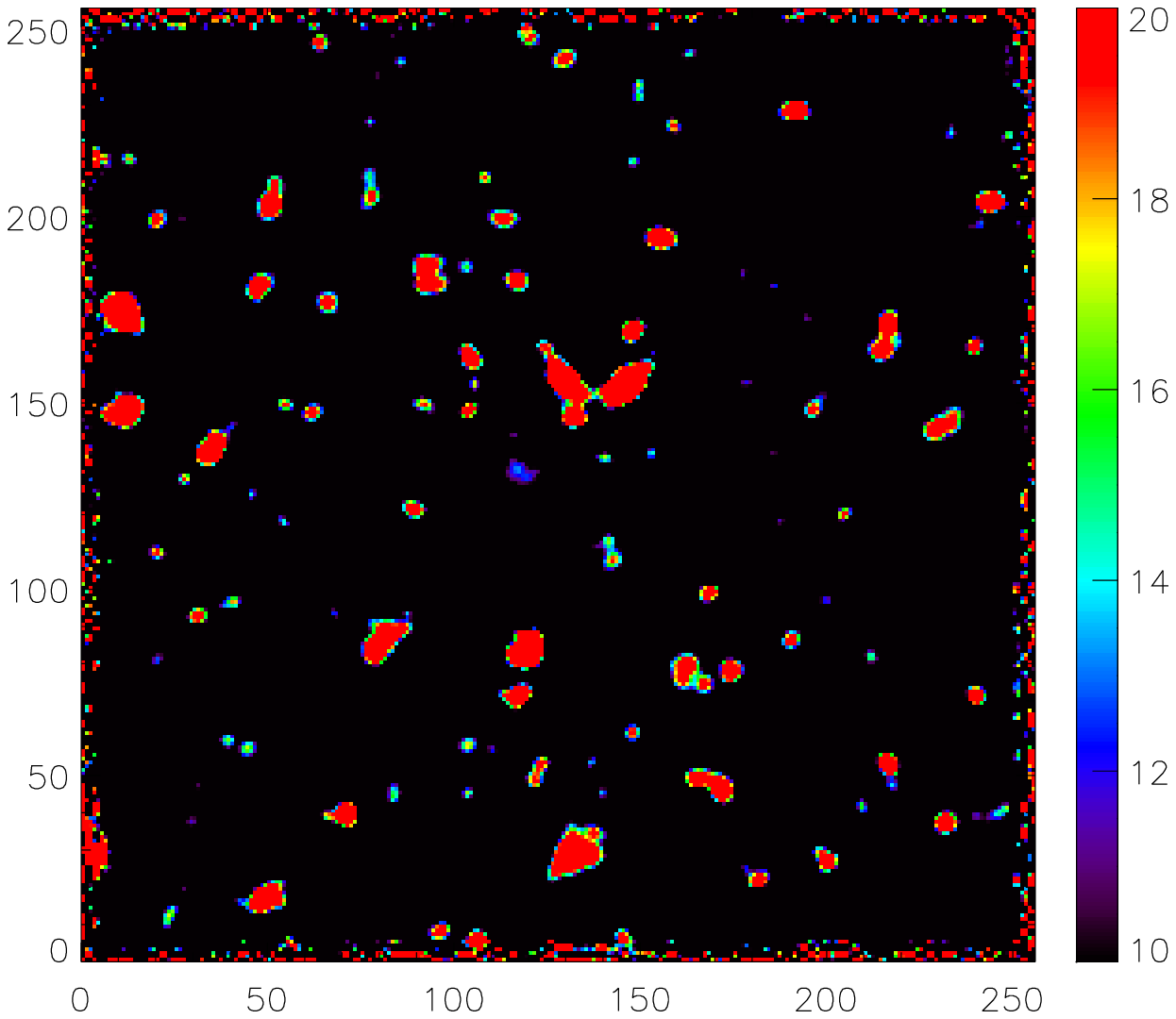}}
\end{center}
\caption{The input SZE (left) and the filtered maps for the  optimal 
filter (center) and the discrete Daubechies wavelet filter (right), 
in an extreme case where the noise level in the maps is varied by a 
factor of several hundred.  The recovered maps shown here are in low 
noise regions, and in this case, the localized nature of the wavelet 
filter allows for a substantial improvement in the reconstruction.  
For realistic noise, this advantage is not detected in our maps.}
\label{fig:varynoise}
\end{figure}

One simple use of a wavelet filter is to account for spatially varying 
noise.  If the statistical properties of the noise is known as a function 
of position, then a wavelet filter can adjust to accommodate this in a 
more natural way than filters with no localized spatial properties.  We 
show an example of this in Figure \ref{fig:varynoise} vs. the optimal 
filter, for an extreme case where the noise fluctuates by a factor of several 
hundred in the maps.  Although the reconstruction is notably improved in 
this highly artificial case, no improvement was evident in our maps 
for more realistic noise.

\end{document}